\documentclass[runningheads]{llncs}
\usepackage{graphicx}
\usepackage{amssymb}
\usepackage{amsmath}
\begin{document}
\title{Automatic Detection and Segmentation of Postoperative Cerebellar Damage Based on Normalization}
%
\titlerunning{Automatic Detection and Segmentation of Postoperative Cerebellar Damage}
%

\author{Silu Zhang \and
Stuart McAfee \and
Zoltan Patay \and
Matthew Scoggins }
\authorrunning{S. Zhang et al.}
%
\institute{Department of Diagnostic Imaging, St. Jude Children's Research Hospital, Memphis, TN 38105
\email{\{silu.zhang,stuart.mcafee,zoltan.patay,matthew.scoggins\}@stjude.org}}


\maketitle              
\begin{abstract}
Surgical resection is a common procedure in the treatment of pediatric posterior fossa tumors. However, surgical damage is often unavoidable and its association with postoperative complications is not well understood. A reliable localization and measure of cerebellar damage is fundamental to study the relationship between the damaged cerebellar regions and postoperative neurological outcomes. Existing cerebellum normalization methods are not reliable on postoperative scans, therefore current approaches to measure surgical damage rely on manual labelling. In this work, we develop a robust algorithm to automatically detect and measure cerebellum damage due to surgery using postoperative 3D T1 magnetic resonance imaging. In our proposed approach, normal brain tissues are first segmented using a Bayesian algorithm customized for postoperative scans. Next, the cerebellum is isolated by nonlinear registration of a whole brain template to the native space. The isolated cerebellum is then normalized into the spatially unbiased atlas (SUIT) space using anatomical information derived from the previous step. Finally, the damage is detected in the atlas space by comparing the normalized cerebellum and the SUIT template. We evaluated our damage detection tool on postoperative scans of 153 patients diagnosed with medulloblastoma based on inspection by human expects. We also designed a simulation to test the proposed approach without human intervention. Our results show that the proposed approach has superior performance on various scenarios. 

\end{abstract}
\section{Introduction}
Surgical resection is a common procedure in the treatment of pediatric posterior fossa tumors. However, surgical resection is associated with potential serious complications~\cite{dubey2009complications}, such as postoperative cerebellar mutism~\cite{van1995transient,gelabert2001mutism,pollack1997posterior,10.1093/neuonc/noaa215.536} and resultant long-term neuropsychological dysfunction~\cite{grill2004critical,levisohn2000neuropsychological}. While brain tissue damage caused by the tumor itself may be minor, it is typically not possible to remove the tumor without impacting surrounding healthy brain tissue. Thus, surgical damage is often unavoidable. Understanding the relationship between the damaged cerebellar regions and postoperative neurological and cognitive outcomes has the potential to guide surgery and significantly affect quality of survival of patients. Therefore, a reliable localization and measure of cerebellar damage is needed and fundamental to conduct this type of investigation.

Defining the volume-of-interest (VOI) for cerebellar damage in the postoperative brain MR image is challenging, even for human experts performing this task manually. First, the damage we want to identify herein is the ``loss of normal brain tissue''. Instead of showing an abnormal intensity in the MR image, this missing tissue presents as just an empty space (cavity) filled with cerebrospinal fluid (CSF). In addition, an abnormal CSF distribution does not necessarily reflect damage, because normal brain tissues may have already been displaced by the presence of the tumor and disease complications (e.g., hydrocephalus) before the surgery. Presence of the tumor near the fourth ventricle complicates this issue, as the ventricle space is connected to the postsurgical cavity. Given these challenges, we aim to develop a fully automatic procedure to define the damage VOI. We measure the damage by comparing the spatially unbiased atlas template (SUIT) of the cerebellum~\cite{diedrichsen2006spatially} and the normalized patient's cerebellum, which distinguishes the missing brain tissue from the fourth ventricle space.

Efforts have been made in cerebellum normalization to understand structure-function relationships~\cite{grosse2021mapping,schmahmann2000mri,diedrichsen2006spatially,diedrichsen2010advances,hernandez2019representative,diedrichsen2009probabilistic}. Grosse et al.~\cite{grosse2021mapping} related cerebellar lesion to the cognitive and motor deficits in pediatric brain tumor survivors. However, their approach to identify cerebellar lesion was achieved manually. A lesion mask was manually created in both native and atlas space. Our approach is fully automatic, and is closely related to SUIT normalization, first proposed in ~\cite{diedrichsen2006spatially} by Diedrichsen. In their approach, an isolation algorithm is applied to segment the cerebellum from the whole brain, using the ICBM152 template and prior information on the brain tissue. The the cerebellum is then normalized to the SUIT atlas via the algorithm proposed in~\cite{ashburner1999nonlinear} and implemented in the SPM2 package~\cite{spm}, an early version of SPM. Diedrichen et al. later developed a 
VOI-based SUIT normalization\cite{diedrichsen2011imaging}, which uses the VOI of the dentate nuclei to guide normalization. The VOI-guided approach yields more accurate normalization, but requires the dentate nuclei VOI as an additional input. The SUIT normalization software is available at~\cite{suit}, which uses SPM12 segmentation for brain tissue segmentation and DARTEL~\cite{ashburner2007fast} as the normalization algorithm in its latest implementation. Despite its success in many applications~\cite{o2008cerebellum,donchin2012cerebellar,kuper2011evidence}, the SUIT normalization technique, however, cannot be directly applied here to detect cerebellar damage, because it does not take the surgical cavity into account and damage to the dentate nuclei may be present. The cerebellum isolation algorithm is likely to fail in the first place by miss-classifying surgical cavity as gray matter, regardless of the normalization algorithm used later. In addition, given an accurate cerebellum mask, the DARTEL algorithm tends to over stretch the brain tissues to match the template, resulting in little or no damage in the atlas space. Therefore, both the brain tissue segmentation and the cerebellum normalization need to be carefully designed for postoperative images.

In this work, we propose a novel automatic approach to detect and measure cerebellar damage. This work has three major contributions: (1) We develop a brain tissue segmentation algorithm that considers the presence of surgical cavity, enlarged ventricles and other non-tissue components such as the presence of absorbable hemostat and blood products, which are commonly seen in postoperative images. 
(2) We develop a customized cerebellum normalization procedure that uses anatomical information in combination with the state-of-the-art nonlinear registration algorithm, SyN~\cite{avants2008symmetric,klein2009evaluation}, implemented in the Advanced Normalization Tools (ANTs)~\cite{ants}. The normalization technique proposed here is more accurate than SUIT normalization on postoperative images, thus can also be useful for applications other than damage detection, such as functional magnetic resonance imaging (fMRI) analyses where mapping from the blood-oxygen-level-dependent (BOLD) signal to the cerebellum functional regions is required.
(3) We provide an automatic and accurate VOI estimate of cerebellar damage in the atlas space, which is vital to study structural damage and function relationship and also to perform group comparisons.

\section{Methods}
\label{sec:method}
\subsection{Overview}
Detecting surgical damage is challenging because the void region in the postoperative image does not necessarily indicate damage. This is evidenced by the following: 1) If the tumor was near the fourth ventricle before resection, the void region of the postoperative image would show both the cerebellar damage and the fourth ventricle space as one unified fluid-filled space. 2) Even for lateral tumors, the volume of the void region does not reflect the volume of damage, because the degree of deformation of normal brain tissues varies by patient. Hence, we detect surgical damage in the atlas space instead of native space. The accuracy of the damage detection then heavily relies on the accuracy of segmentation and normalization algorithms. Therefore, we specifically design a brain tissue segmentation algorithm and a cerebellum normalization approach to address this issue, which constitute two major components of the overall damage detection tool.

\begin{figure}
\includegraphics[width=\textwidth]{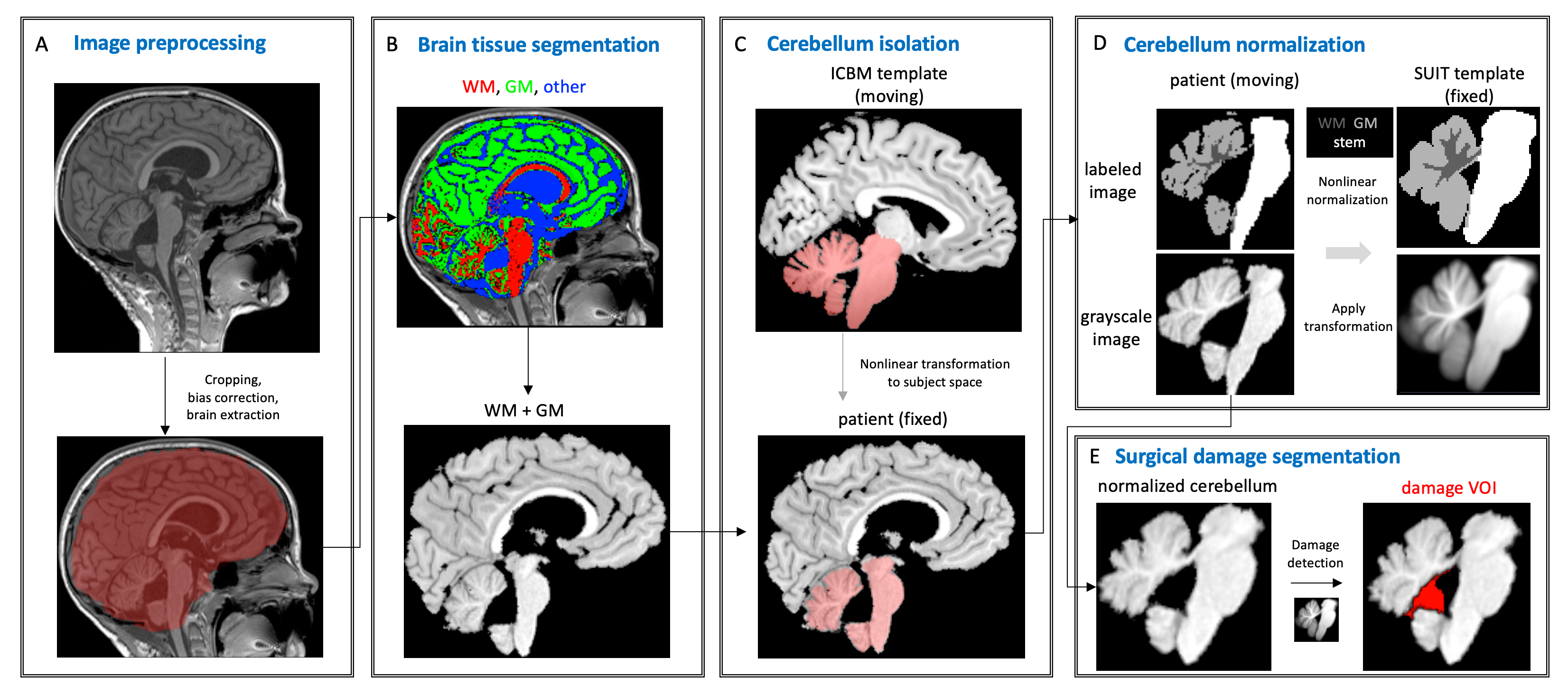}
\caption{An overview of the proposed framework for detecting cerebellar surgical damage.} \label{fig:workflow}
\end{figure}

An overview of the proposed framework is shown in Fig.~\ref{fig:workflow}. The 3D T1 image is first preprocessed by cropping, bias correction and brain extraction. The white matter (WM) and gray matter (GM) are then segmented via a brain tissue segmentation algorithm we developed taking the surgical cavity into account (see detailed description in Section~\ref{sec:brain_tissue_seg}). The cerebellum is isolated by nonlinear registration of the ICBM template to the native space, keeping only WM and GM in the moving and fixed image. The isolated cerebellum of the patient is then normalized into the SUIT template space~\cite{diedrichsen2006spatially} using labeled (background: 0, WM: 1, GM: 2, stem: 3) fixed and moving images. Finally, the surgical damage is detected by comparing the difference between the normalized image and the SUIT template.
\subsection{Templates and processing}
\label{sec:templates}
\subsubsection{ICBM} The ICBM template is available at UCLA brain mapping center~\cite{icbmtempalate} and is used here as our the whole brain template. Our implementation uses only WM and GM for more accurate registration to the native space. WM and GM can be obtained by thresholding the template after brain extraction using FSL BET~\cite{smith2002fast}. The cerebellum mask of ICBM is the combined region of the zones with the following labels: 58, 67, 237, 238, and 251. The probability maps of WM and GM of this atlas~\cite{icbmprob} are used in our brain tissue segmentation algorithm as prior probabilities.
\subsubsection{SUIT}
The SUIT atlas~\cite{diedrichsen2006spatially} is used as the cerebellum template for cerebellum normalization. The same brain tissue segmentation algorithm is used to generate probability maps of WM and GM and label this template (WM: 1, GM: 2). The stem VOI was created manually and labeled as 3.

\begin{figure}
\centering
\includegraphics[width=3.5in]{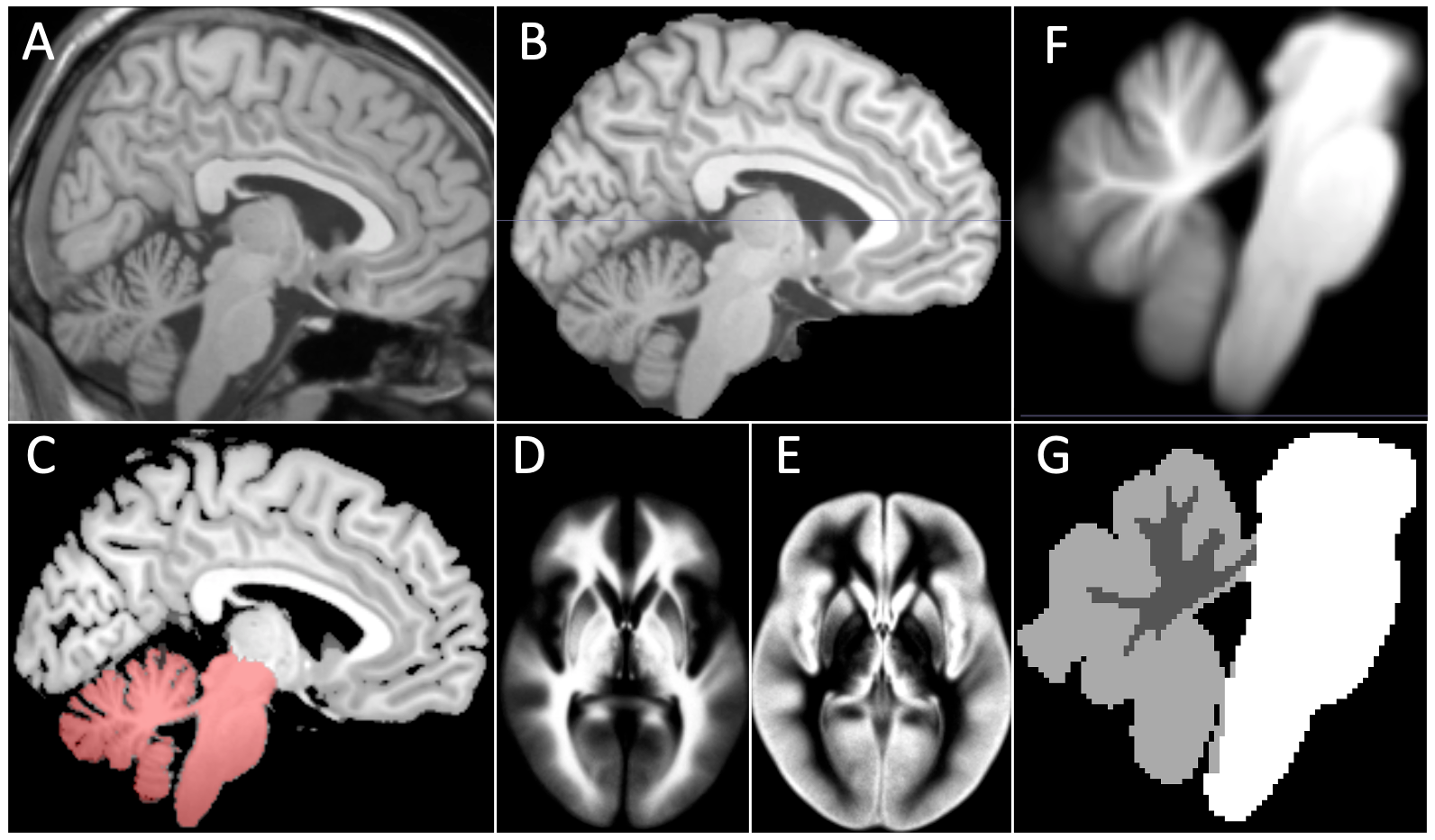}
\caption{Original and processed templates used in this study. The ICBM atlas (A-E). A: original (sagittal); B: brain extracted (sagittal); C: cerebellum mask overlaid on the thresholded template (sagittal); D: white matter probability map (axial); E: gray matter probability map (axial).  The SUIT atlas (F-G). F: original (sagittal); G: labeled (sagittal)} \label{fig:templates}
\end{figure}

The original and processed templates used in our implementation are shown in Fig~\ref{fig:templates}.
\subsection{Input preprocessing}
\label{sec:preprocessing}
Neck and lower head are first removed using FSL robustfov\cite{robustfov} from the input postsugical 3D T1 image. The cropped image is then bias corrected using N4 bias correction~\cite{tustison2010n4itk}, and brain extracted using FSL BET~\cite{smith2002fast}. Because of the presence of the surgical cavity and potentially contaminated brain tissues, the BET mask is generally not satisfactory. Thus, a better brain mask is obtained by affine registration of the brain extracted ICBM template to the brain extracted input image, and then binarizing the warped template. 
\subsection{Brain tissue segmentation}
\label{sec:brain_tissue_seg}
Patients being treated for posterior fossa brain tumors present with abnormalities that complicate standard neuroimaging anatomical analysis such as segmentation of brain tissues.  In addition to the presence of a tumor, or relatively large postsurgical cavity (which may progressively contract later), abnormalities such as enlarged ventricles (i.e., hydrocephalus) are common. Therefore, commonly used brain tissue segmentation tools (e.g., SPM unified segmentation~\cite{ashburner2005unified}) designed for the healthy population are susceptible to failure on postoperative images. Here we present a brain tissue segmentation approach modified from the brain tumor segmentation algorithm based on outlier detection, proposed in ~\cite{prastawa2004brain}. As this work is aimed to detect damage on postoperative images, we assume that there is no tumor present in the image. Therefore, only three classes are modeled in our algorithm: WM, GM and other, i.e., class label $\mathbb{Y}=\{\mathrm{WM},\mathrm{GM},\mathrm{other}\}$. The ``other'' class contains primarily CSF, but may also contain blood products, hemostat, non-brain regions (due to imperfect brain extraction), and other components that are not considered as normal brain tissue. 

The WM and GM probabilities are initialized as ICBM atlas priors (affine transformed into the native space). Because of the presence of surgical cavity, enlarged ventricles and other non-tissue components, probability of the ``other'' class is initialized as uniform distribution ($
\Pr(\mathrm{other})=0.5$), rather than using the CSF prior of ICBM.  Then, the probability density function of each class, denoted as $p(x|Y)$, is estimated via kernel density estimation using voxels sampled according to prior probability. Because the WM and GM samples can have contaminants (due to inaccurate priors), an outlier detection approach is used to detect and remove the contaminants (key idea proposed in ~\cite{prastawa2004brain}). Outlier detection is performed by using the Minimum Covariance Determinant estimator, setting the support fraction to 0.5. Unlike ~\cite{prastawa2004brain} where T1 and T2 are taken as inputs, our implementation uses only 3D T1 as the input, as it provides good spatial resolution and good contrast between WM, GM, and others. The outlier detection in our implementation is hence performed in one dimension (T1 only) instead of two (T1 and T2 as in ~\cite{prastawa2004brain}), and not performed for the ``other'' class, since we allow this class to have contaminants. The posterior probabilities are then calculated according to Bayes’ theorem:
$$
Pr(Y|x)=\frac{p(x|Y)Pr(Y)}{p(x)},
$$
where $p(x)=\sum_{Y\in \mathbb{Y}}p(x|Y)Pr(Y).$
The posteriors are used as priors in the next iteration, and the probabilities are updated through a total of 3 iterations in our implementation. The final WM and GM masks are generated by thresholding the smoothed posteriors at 0.5. The union mask of WM and GM is denoted as $\mathrm{WM} \cup \mathrm{GM}$.

\subsection{Cerebellum isolation, normalization and damage detection}
\subsubsection{Cerebellum isolation}
The cerebellum is isolated by non-linear transformation of the ICBM atlas to the native space. To achieve a better alignment, only WM and GM regions of the moving and the fixed images are used for registration. Registration is performed using ANTs~\cite{ants}, with affine transformation as the initial step followed by nonlinear transformation using the SyN algorithm~\cite{avants2008symmetric}. The cerebellum mask of the ICBM template is then warped into the native space. The warped ICBM cerebellum mask is denoted as $\mathrm{Cb_w}$. The patient's cerebellum mask is the intersection of $\mathrm{WM} \cup \mathrm{GM}$ and $\mathrm{Cb_w}$ adjusted by morphological operations and smoothing.
\subsubsection{Cerebellum normalization}
The initial step of cerebellum normalization is affine registration of the patient's cerebellum (along with the WM and GM masks) to the SUIT template. Next, to achieve detailed matching in the nonlinear transformation, both the moving (affine transformed input image) and fixed (SUIT template) images are labeled (background: 0, WM: 1, GM: 2, stem: 3). The stem mask of the patient was generated after affine registration using the template's label information. The ANTs SyN is used to perform the final step of normalization. 
\subsubsection{Cerebellar damage detection}
We define the cerebellar damage as the missing brain tissue (WM or GM) of the patient's cerebellum. Therefore, assuming a perfect normalization, the damage VOI is segmented in the SUIT atlas space by subtracting the normalized cerebellum from the template after binarization of both, followed by taking the largest connected component. Detecting damage in the atlas space has two major advantages: 1) the fourth ventricle space is not counted as damage because it has the background intensity in the SUIT template; 2) the damage VOIs can be compared (e.g., size, location) across patients.
\section{Experimental Results}
\subsection{Datasets}
We used 3D T1 scans of a prospective study (SJMB12; NCT 01878617) at St. Jude Children's Research Hospital to test our cerebellar damage detection algorithm. Specifically, we included 245 patients diagnosed with medulloblastoma (MB) and had the complete tumor resected (no remaining tumor or metastasis). Our damage detection approach was applied on one postoperative 3D T1 scan for each patient. 3D T1 images were acquired 1-368 days after the last surgery. Evaluation of performance was established on a subset of imaging (N=153) with ground truth created by human experts. To better understand the performance of the proposed approach in wider scenarios, we also created a simulated dataset by generating damages on healthy brain images. The manual evaluation procedure and the simulation design are described below.

\subsubsection{The CMS Dataset}
Manual evaluation was performed on results of 153 patients who either had postoperative cerebellar mutism syndrome (CMS) or was asymptomatic after surgery, for the purpose of studying CMS related surgical damage.  Evaluation of the results is challenging, as it is extremely difficult for human experts to draw the damage VOI on the SUIT template by just looking at the 3D T1 in native space, blinded to the auto generated results (the normalized T1 and auto mask). To reduce bias, we asked two raters, an expert in cerebellar anatomy and a neuroradiologist, to evaluate the damage VOIs generated by the algorithm. During the evaluation, the raters first inspected the 3D T1 scan, and then examined the normalized image and the damage mask in the atlas space. Each damage mask was either considered accurate as is or edited manually and approved by both raters. Then, the performance of the algorithm was measured by the S$\phi$rensen-Dice Similarity Coefficient (Dice) between the auto generated mask (AM) and the approved ground truth (GT) mask. The Dice score is defined as
$$
Dice = \frac{2|GT\cap AM|}{|GT|+|AM|}.
$$
\subsubsection{Simulated Dataset}
The evaluation on the CMS dataset had some limitations. First, this dataset was relatively small, and only 13 out of 153 cases had a lateral damage. Thus, it's hard to conclude how well the algorithm does in detecting lateral damages. Second, the raters were potentially biased by the auto prediction due the natural of the evaluation process. Hence, we designed a simulation procedure to generate sufficient data that covers both ventricular and lateral damages of different sizes.

The main idea of the simulation was to generate a postoperative 3D T1 image in the native space, the normalized cerebellum in the SUIT atlas space, a damage VOI showing the missing tissue, and the transformation that maps the two spaces. To achieve this, we also used 3D T1 images of 69 healthy children (not necessarily age matched to the patients) recruited for a separate study. The healthy controls do not have a history of cancer or are a first degree relative or direct friend of the MB patients. The simulation contains the following steps: 1) Brain tissue segmentation and cerebellum isolation (described in Section~\ref{sec:method}) was performed on a 3D T1 of a healthy brain. 2) The SUIT template was registered to the isolated healthy cerebellum using ATNs SyN. The transformation of this registration was reliable because the anatomies of both the moving and fixed images were complete. The registration was also inspected to ensure quality. 3) A simulated damage VOI was created on the SUIT atlas by randomly sampling one VOI from all the 245 VOIs generated by the algorithm on the SJMB12 patients (including those that were not verified by human expects), followed by applying random deformation. 4) The simulated damage VOI was then overlaid on the native image by applying the transformation generated from step 2. 5) The simulated postoperative 3D T1 image was created by replacing the damage VOI in the native space with CSF intensities using Gaussian distribution $\mathcal{N}(\mu,10)$, followed by random deformation of the whole image. The mean of CSF $\mu$ can be set as the median intensity in the ``other'' mask from brain tissue segmentation, since the ``other'' class has no containment in healthy brains. 

We generated 10 random damage VOIs for each of the 69 healthy children, which resulted in a total of 690 pairs of simulated 3D T1 and GT damage VOI. Our damage detection tool was applied on the simulated 3D T1 and the auto generated damage VOI was compared to GT using Dice score.
\subsection{Results}
\subsubsection{The CMS Dataset}
The average run time of the whole damage detection pipeline was about 40 minutes and the average memory usage was about 2 GB. 
82 out of 153 damage masks were considered accurate as is, and for the remaining cases that required manual adjustment, the average time needed for editing was about 10 minutes. The mean Dice score of 153 patients was $0.892 (\pm 0.217)$. The GT mask size range from 0 (no damage) to 37210 $mm^3$, with a mean of 3455 $mm^3$. In this dataset, there were 4 patients that had no surigical damage observed by the raters. Because the algorithm always detect the missing voxels in the normalized image as damage, the Dice score for these 4 cases was 0. The masks that were approved as is had a Dice score of 1. We observed that the performance of the algorithm depends on the size of the damage. The smaller the damage, the harder to detect. This is because when the surgical cavity is small, the result can be skewed by MR artifact, abnormal anatomy not caused by resection, or misalignment during normalization, etc. The mean Dice scores of different damage size ranges are summarized in Table 1.
\begin{table}
\caption{Performance (Dice) of the cerebellar damage detection tool on patients with different damage sizes.} \label{table:dice_size}
\includegraphics[width=\textwidth]{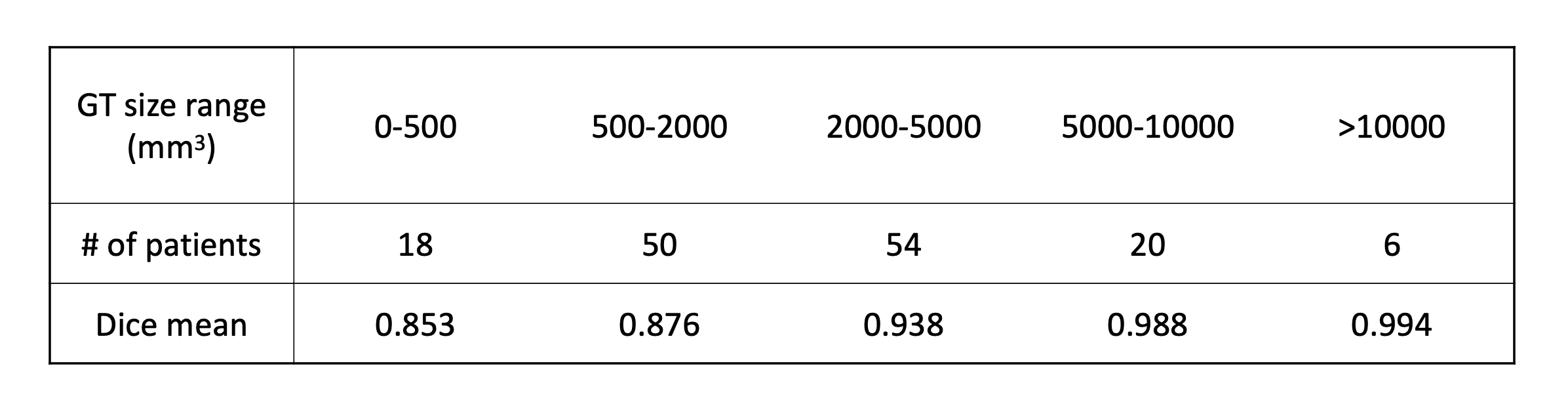}
\end{table}

\begin{figure}
\includegraphics[width=\textwidth]{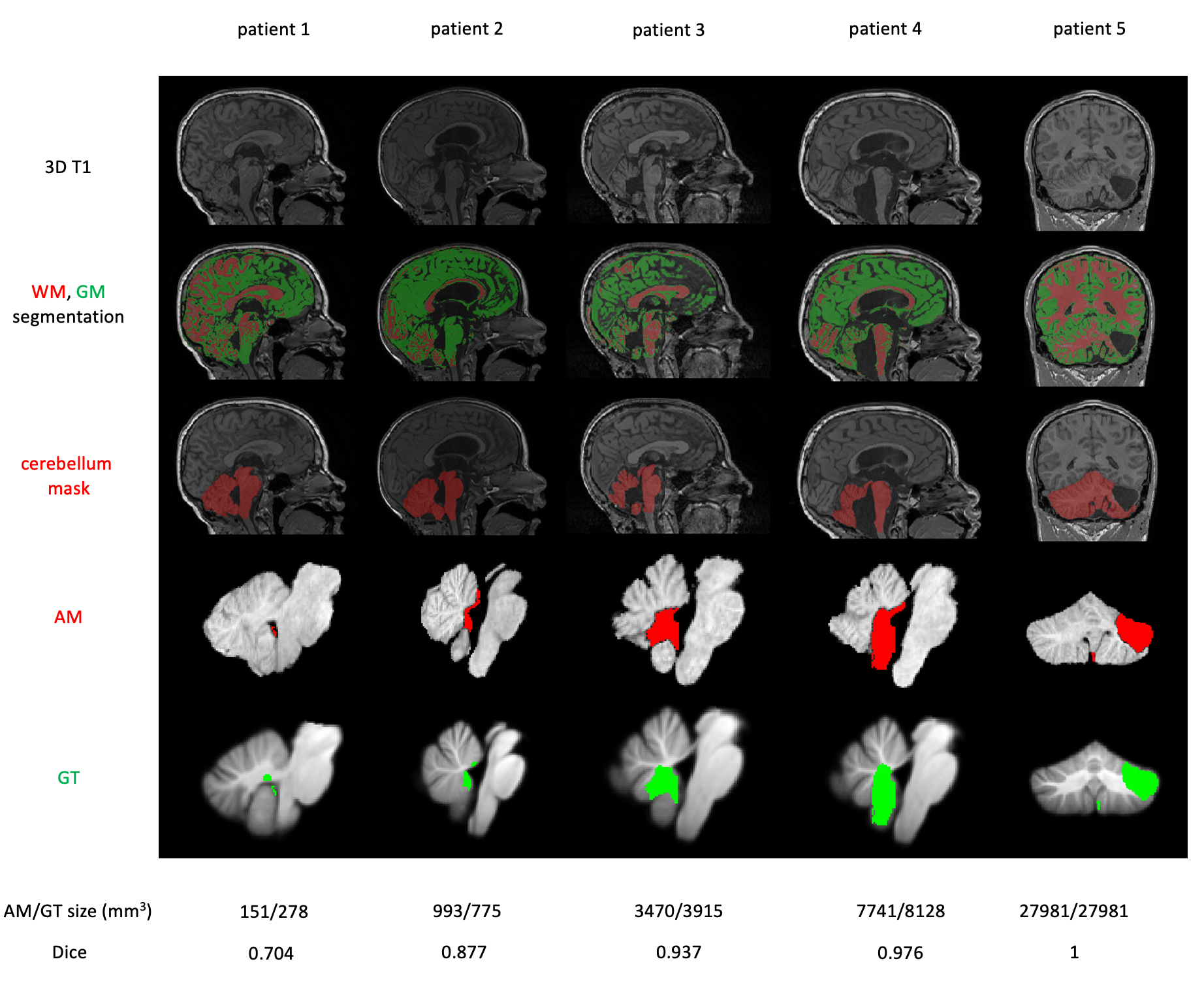}
\caption{Representative examples of patients with different cerebellar damages in size and location. First row: cropped 3D T1. Second row: white matter (WM, red) and gray matter (GM, green) segmentations. Third row: cerebellum mask (red) overlaid on 3D T1. Fourth row: auto generated damage mask (AM, red) overlaid on the normalized cerebellum. Fifth row: approved ground truth (GT, green) mask overlaid on the SUIT template. Patient 1-4 had a damage near the fourth ventricle and results are shown in the sagittal plane. Patient 5 had a lateral damage and results are shown in the coronal plane.} \label{fig:results}
\end{figure}

To better demonstrate our results, we selected one patient from each size range that had a Dice score close to the mean value (per Table~\ref{table:dice_size}). The results of the five example cases are shown in Fig.~\ref{fig:results}. 
The first row shows the 3D T1 scans that the algorithm took as inputs. The second row shows the WM (red) and GM (green) segmentations from our brain tissue segmentation model. The third row shows the cerebellar mask (red) overlaid on the 3D T1 image. The fourth row shows the auto generated cerebellar damage mask (red) overlaid on the normalized cerebellum in the atlas space. The fifth row shows the GT mask (green) overlaid on the SUIT template. The results show the superior performance of our proposed approach under various scenarios (different damage sizes and locations). 

\begin{figure}
\includegraphics[width=\textwidth]{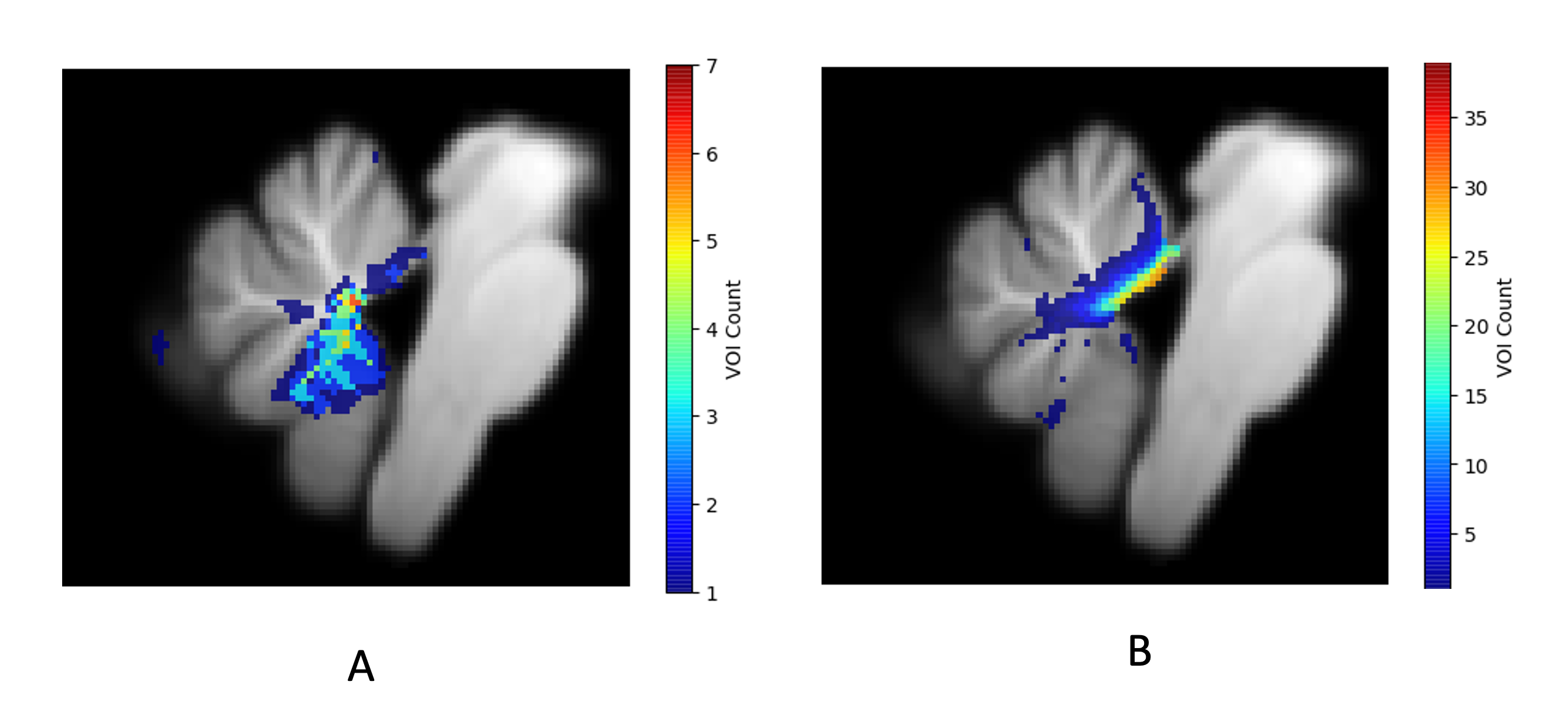}
\caption{The heatmap of false negative (A) and false positive (B) voxels detected by the proposed damage detection approach.} \label{fig:FN_FP}
\end{figure}

Figure~\ref{fig:FN_FP} shows the heatmaps of false negative (FN) (Figure~\ref{fig:FN_FP}A) and false positive (FP) (Figure~\ref{fig:FN_FP}B) voxels detected by our approach. The count of FN is relatively small, with the maximum count being 7, and FN voxels appear to be randomly distributed. The FP voxels are centered near the superior cerebellar peduncles (SCPs). This is a result from misalignment in the SCP region during the normalization process. The SCPs have very thin structure and therefore are under emphasized by the registration algorithm.

\subsubsection{Simulated Dataset}
\begin{figure}
\includegraphics[width=\textwidth]{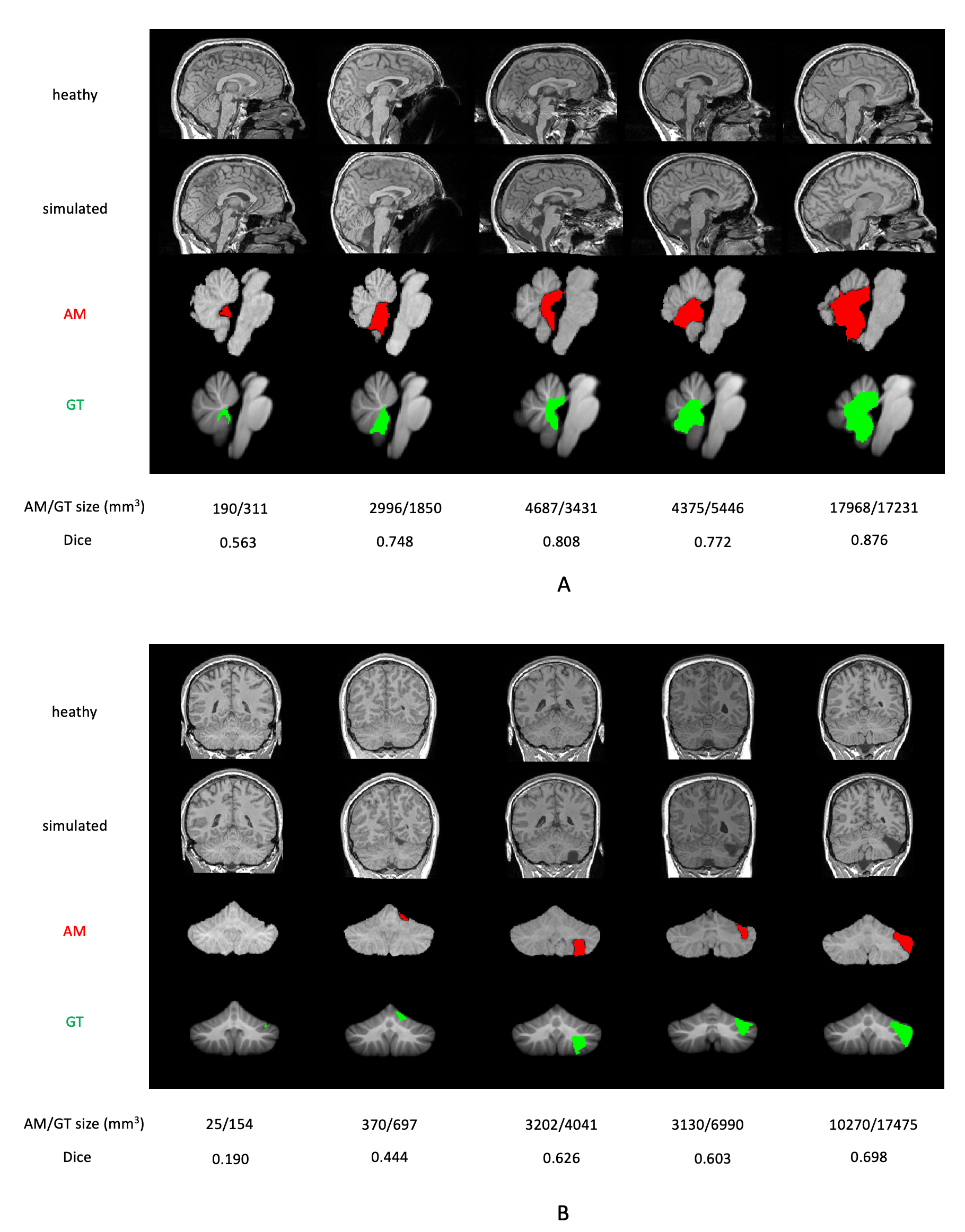}
\caption{Visualization of representative examples of the simulated data and damage detection results. First row: the healthy 3D T1 brain image used for simulation. Second row: the simulated postoperative 3D T1 image after applying the simulated damage VOI. Third row: the automatically detected damage mask (AM, red) on the normalized cerebellum. Fourth row: the ground truth (GT, green) simulated damage mask on the SUIT template. A) Ventricular damages. Results are shown in the saggittal plane. B) Lateral damages. Results are shown in the coronal plane.} \label{fig:simulation}
\end{figure}

The performance of the damage detection algorithm on the simulated data is summarized in Table~\ref{table:simulation}. The mean Dice score on ventricular damages was 0.723, higher than the score on lateral cases, which was 0.365. Similar to what was observed from the CMS dataset, the algorithm was more accurate on damages with larger size. Qualitative results are shown in Fig.~\ref{fig:simulation}. For each size range, we selected a representative example that had the closest Dice to the mean value (per Table~\ref{table:simulation}) for both ventricular (Fig.~\ref{fig:simulation}A) and lateral damages (Fig.~\ref{fig:simulation}B). The algorithm was able to accurately locate the damage, even when damage was small (first case in Fig.~\ref{fig:simulation}A and first two cases in Fig.~\ref{fig:simulation}B). Because the normalization was not perfect, a small degree of misalignment of the healthy brain tissues near the damage would dramatically lower the dice score if the missing VOI was small. These results suggest that inspection and manual editing by human expect is required to ensure the quality of damage detection. Overall, the simulation provided an unbiased evaluation of the proposed damage detection approach, which achieved reasonable normalization of the cerebellum and robust damage detection.
\begin{table}
\caption{Performance (Dice) of the cerebellar damage detection tool on simulated datasets categorized by size and location.} \label{table:simulation}
\includegraphics[width=\textwidth]{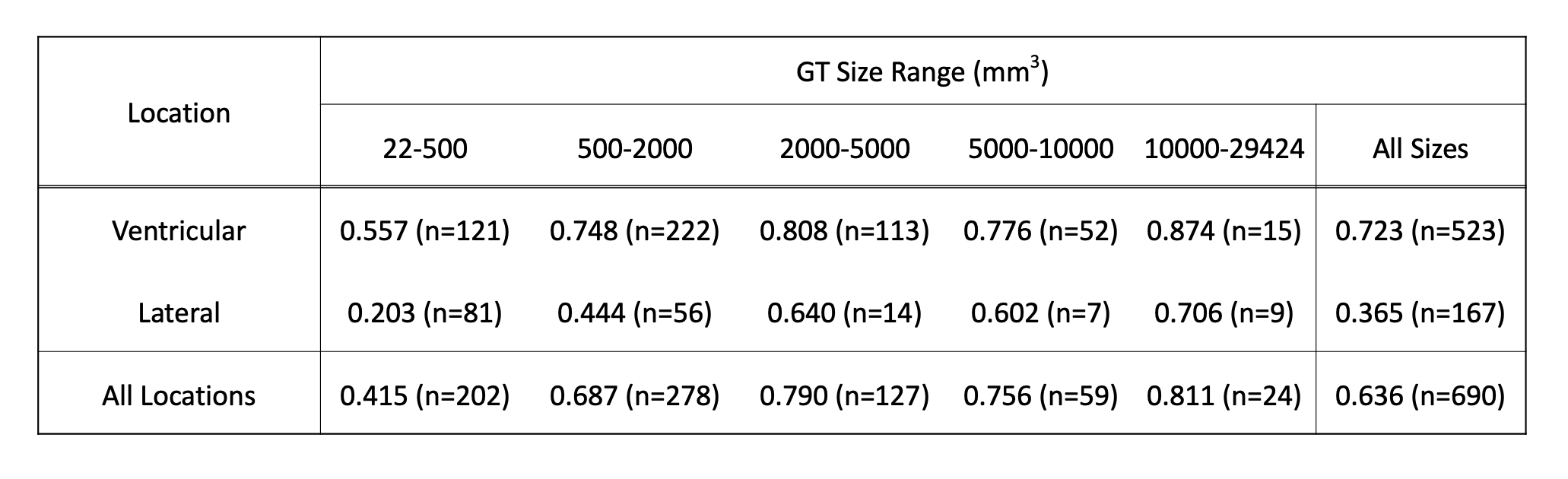}
\end{table}

\section{Discussions}
Existing cerebellum normalization tools are typically designed for healthy brain imaging data, thus tend to fail when lesions are present. As a by-product of this cerebellar damage detection approach, the normalization of cerebellum is also suitable for broader neuroimaging applications (e.g., fMRI preprocessing and analysis), and is more accurate than SUIT normalization~\cite{diedrichsen2011imaging} on postoperative images. Our normalization is more accurate because (1) it has a more accurate cerebellum mask and WM/GM segmentation to start with, (2) labeled inputs allow for easier and better optimization, and (3) the choice of state-of-the art nonlinear registration algorithm, Syn~\cite{klein2009evaluation}. 

Our work has some limitations: (1) An important assumption of this approach is the absence of the tumor in the image. Hence, this method cannot handle cases with partial tumor resection or metastases. (2) Our brain tissue segmentation algorithm assumes everything with abnormal intensity (hyper or hypo) is not healthy brain tissue, which may not always be the case. For example, brain tissue might be stained by blood and therefore have an abnormal appearance in the image, but be otherwise healthy. However, this is not easy to tell from image at the time point immediately following surgery. MR scans at later time points may further inform healthy or damaged anatomy after the stain has been removed via natural processes. (3) Our damage detection algorithm is not learning-based. As a result, its performance can not be improved by collecting more accurate labels (manually edited masks). Our future work is to develop a deep learning based model to further improve the performance by taking advantage of the inputs from neuroradiologists.

\section{Conclusions}
In this paper, we present a fully automatic localization and measurement of postoperative damage in the cerebellum in the atlas space using postoperative MRI. The proposed framework has a novel brain tissue segmentation algorithms that considers surgery-related features (e.g., postsurgical cavity, absorbable hemostat, blood products) and other non-surgical features (e.g., enlarged ventricles), a more accurate cerebellum normalization procedure, and provides quantification of cerebellar damage in the atlas space. This automation greatly reduced human annotation time from hours to about 10 minutes. The normalization proposed in this work is also helpful to perform other neuroimaging analyses (e.g., fMRI analyses) where cerebellum normalization is needed, especially on postoperative images.

%
%
%
\bibliographystyle{splncs04}
\bibliography{references}





\end{document}